\documentstyle[12pt]{article}
\evensidemargin \oddsidemargin
\def\ds{\displaystyle}
\def\beq{\begin{equation}}
\def\eeq{\end{equation}}
\def\bea{\begin{eqnarray}}
\def\eea{\end{eqnarray}}
\def\ba{\begin{array}}
\def\ea{\end{array}}
\def\del{\partial}

\def\t{\widetilde}
\def\G{\Gamma}
\def\O{\Omega}
\def\L{\Lambda}
\def\e{\epsilon}
\begin{document}
\pagestyle{empty}
\begin{flushright}
hep-th/9912236\\
CPHT-S756.1299\\
\end{flushright}
\vspace*{.5cm}
\begin{center}
\vspace{.5cm}
{\Large\bf SO(d,d) Transformations of Ramond-Ramond Fields}\\
\vspace{.5cm}
{\Large\bf and Space-time Spinors}\\
\vspace*{1cm}
{\bf S. F. Hassan}{\footnote{\tt e-mail: fawad@cpht.polytechnique.fr}} \\
\vspace{.2cm}
{\it Centre de Physique Th\'{e}orique, Ecole Polytechnique, \\
91128 Palaiseau, France}\\ 
\vspace{1cm}
{\bf Abstract}
\begin{quote}
We explicitly construct the $SO(d,d)$ transformations of Ramond-Ramond
field strengths and potentials, along with those of the space-time
supersymmetry parameters, the gravitinos and the dilatinos in type-II
theories. The results include the case when the $SO(d,d)$
transformation involves the time direction. The derivation is based on
the compatibility of $SO(d,d)$ transformations with space-time
supersymmetry, which automatically guarantees compatibility with the
equations of motion. It involves constructing the spinor
representation of a twist that an $SO(d,d)$ action induces between the
local Lorentz frames associated with the left- and right-moving
sectors of the worldsheet theory. The relation to the transformation
of R-R potentials as SO(d,d) spinors is also clarified.    
\end{quote}
\end{center}
\vfill
\begin{flushleft}
{Dec 1999}\\
\vspace{.5cm}
\end{flushleft}
\newpage
\setcounter{footnote}{0}
\pagestyle{plain}
\section{Introduction}
The $O(d,d)$ transformations in string theory, also referred to as the
{\it Narain group} or the {\it generalized T-duality group}, have
played an important role in understanding toroidal compactifications
and dualities, as well as in constructing classical solutions to the
low-energy equations of motion \cite{KSN,GRV,MV,SEN,HS,MS}. The action of
the $O(d,d)$ group on the NS-NS sector fields (the graviton, the
antisymmetric tensor field and the dilaton), which are all assumed to
be independent of $d$ coordinates, has been known for a long time.
In the worldsheet formalism, the $O(d,d)$ action on these fields can
be obtained either by a canonical transformation \cite{GRV,MV}, 
or by gauging appropriate isometries in a non-linear $\sigma$-model
\cite{GR}. In the framework of the low-energy effective theory, one can
derive the transformation of the fields by writing the effective
action for the NS-NS fields in a manifestly $O(d,d)$ invariant form
\cite{MV,SEN,HS,MS}.    

Deriving the $O(d,d)$ transformation of the Ramond-Ramond fields has
turned out to be more complicated. Since the fundamental string does
not carry Ramond-Ramond charges, the usual NSR formalism worldsheet
methods cannot be used. In \cite{BHO} the equations of motion in IIA
and IIB supergravities were used to derive the transformation of the
Ramond-Ramond potentials under the single T-duality subgroup of
$O(d,d)$ which interchanges the two theories. However, the
generalization of this approach to the full $O(d,d)$ group (or to its
$SO(d,d)$ part, if we do not want to interchange IIA and IIB theories)
has not been straightforward. Before describing our derivation of the
transformation, which is based on supersymmetry, let us briefly review
an interesting, though as yet inconclusive, alternative approach that
has so far been followed. 

In \cite{HT} it was observed that when type-II theories are
compactified on $T^6$, the resulting R-R scalars fill up a
Majorana-Weyl spinor representation of the associated T-duality group
$SO(6,6)$. This observation was further generalized in \cite{EW} where
it was pointed out that the components of R-R potentials in type-II
theories compactified on $T^d$, when arranged in representations of
the Lorentz group $SO(9-d,1)$ of the uncompactified space, fill up
Majorana-Weyl spinor representations of the $SO(d,d)$ group. This
observation is mainly based on the decomposition of representations of
the U-duality group of the compactified theory in terms of the
representations of its S-duality and T-duality subgroups. The
U-duality group contains NS-NS and R-R charges in the same multiplet
and, after this decomposition, the R-R charges are found to fall in
the spinor representations of the T-duality group $SO(d,d)$. In
\cite{AAFFT}, this problem was studied in more detail from the group
theory perspective. The explicit construction of this spinor
representation in terms of the supergravity fields was undertaken in
\cite{BMZ} for a specific case of $T^3$ compactification of type-IIA
theory, and was generalized to $T^d$ compactifications in \cite{FOT}.
In this approach, one constructs $SO(d,d)$ Majorana-Weyl spinors out
of combinations of R-R potentials and the NS-NS $2$-form field. The
guiding principle in this construction is the requirement that, in
terms of these spinors, the kinetic energy terms for the R-R fields in
the low-energy effective action should have a manifestly $SO(d,d)$
invariant from. While, as such, the construction ignores the
electric-magnetic duality constraints between R-R field strengths
(including the self-duality of the type-IIB $5$-form), nevertheless,
one expects it to be consistent with these constraints and to be
extendible to include the R-R dependent Chern-Simons terms in the
low-energy effective action. Although this {\it formally} proves the
$SO(d,d)$ invariance of the R-R kinetic energy terms in the low-energy
effective action, so far it has not been possible to extract from this
the transformation of R-R potentials and field strengths under generic
non-trivial $SO(d,d)$ transformations (The reason for this difficulty
will be clarified in section 6). It should be mentioned that in the
case of IIA on $T^3$, the transformation of the R-R scalar (in 7
dimensions) as an $SO(3,3)$ spinor has also been studied in the
framework non-commutative super Yang-Mills compactifications of Matrix
theory \cite{BMZ, KS}. However, these results are not valid in string
theory except in the matrix theory limit of $G_{ij}\rightarrow 0$,
where many terms in the transformation drop out.

In this paper, we determine the $SO(d,d)$ transformations of the
Ramond-Ramond field strengths and potentials using a more
straightforward approach. In the process we also determine the
$SO(d,d)$ transformations of the space-time supersymmetry parameters
and R-NS fields (the gravitinos and the dilatinos). Our approach is
based on the compatibility of $SO(d,d)$ transformations with
space-time supersymmetry. As a result, our derivation automatically
guarantees the $SO(d,d)$ covariance of the equations of motion (and
hence, the invariance of the associated low-energy effective actions
including the Chern-Simons terms), and is consistent with the
electric-magnetic duality constraints on R-R field strengths, since in
type-II theories all these are determined by supersymmetry. The method
we use is a generalization of the one used in \cite{SFH4} for a single
T-duality transformation and can be summarized as follows. In type-II
supergravity theories, every spinor index originates either in the
left-moving or in the right-moving worldsheet sector of the underlying
string theory. On the other hand, it has been known that in flat space
non-trivial $O(d,d)$ transformations correspond to transforming the
left-moving and right-moving parts of the space-time coordinates by
independent Lorentz rotations; $\del_+ X\rightarrow R\,\del_+ X$,
$\del_- X\rightarrow S\,\del_- X$ \cite{SEN,HS}. This feature also
survives in curved backgrounds where, ignoring a contribution from the
worldsheet fermions, one gets $\del_\pm X\rightarrow Q_\pm\,\del_\pm
X$, though now $Q_\pm$ are space-time dependent through their
dependences on the background fields \cite{BS,SFH3}. Thus, one can
regard a non-trivial $O(d,d)$ transformation as twisting the left- and
the right-moving sectors of the worldsheet theory with respect to each
other. By supersymmetry, this translates to transforming the spinor
indices originating in the two worldsheet sectors by two different
$O(d,d)$ induced local Lorentz transformations. In particular, we may
choose a convention in which the spinor index associated with, say,
the right-moving sector remains unchanged, and only the one associated
with the left-moving sector transforms. This determines the basic
$O(d,d)$ action on the space-time spinors as well as on the
Ramond-Ramond fields which can be combined into bispinors. The exact
form of the transformation is fixed by the $O(d,d)$ covariance of the
supersymmetry transformations. To make this structure behind the
transformation of the R-R fields manifest, it is important that the
supersymmetry transformations are written in terms of variables that
appear more natural in string theory. Note that in this approach we
only have to construct the spinor representation of the local Lorentz
twist induced by $O(d,d)$ which is an element of a subgroup of the
Lorentz group $O(9,1)$, and not that of the full $O(d,d)$ group. The
explicit construction we give here is for the $SO(d,d)$ part of the
group. The single T-duality case has been discusses in detail in
\cite{SFH4}. 

This paper is organized as follows: In section 2 we describe our
notation and briefly review some aspects of the $O(d,d)$
transformations that will be needed in the rest of the paper. In
particular, we emphasize that after a non-trivial O(d,d)
transformation, the transformed theory contains two different
vielbeins that are related by a local Lorentz transformation. This is
interpreted as a twist between the local Lorentz frames associated
with the left- and right-moving sectors of the worldsheet theory. In
section 3 we argue that this local Lorentz twist should be absorbed by
the spinors and hence construct its spinor representation. We discuss
the flat-space T-duality limit in some detail to highlight some
important features of the spinor representation. In section 4, we
obtain the $O(d,d)$ action on the supersymmetry variation parameters,
the gravitinos and the dilatinos. In section 5, we use the
transformation of the space-time spinors along with the explicit
construction of the spinor representation of the local Lorentz twist
to obtain the $SO(d,d)$ action on the Ramond-Ramond field strengths.
Then by a simple argument we show that the transformations of the R-R
potentials have exactly the same form as those of the field strengths.
After restricting to type-IIA on $T^3$, and taking the
$G_{ij}\rightarrow 0$ limit, we reproduce the results obtained in the
NCSYM formalism. We also write a simple $SO(d,d)$ covariant form for
the R-R kinetic energy terms in the low-energy effective action. In
section 6, we discuss the relation between our approach and the
alternative approach in which components of R-R potentials combine 
into $SO(d,d)$ spinors. Our conclusions are summarized in section 7.
Appendix A outlines the steps for the derivation of two sets of
equations needed in section 4, and Appendix B contains some
$\G$-matrix results. 

\section{Preliminaries}
In this section, we will describe our notation and conventions and  
then briefly recall the action of the $SO(d,d)$ group on the massless
NS-NS sector fields for later reference. Everything in this section
applies equally well to the $O(d,d)$ group but we restrict ourselves
to elements with positive determinant which do not interchange
type-IIA and type-IIB theories. 
 
\noindent\underline{{\it Notation} :} We denote the 10-dimensional
space-time coordinates by $X^M$ ($M=0,\dots, 9$), and assume that all
fields are independent of the $d$ coordinates $X^i$, for $i=0,\cdots
d-1$. For the sake of generality, we have assumed that these
coordinates also include time. The remaining $10-d$ coordinates are
denoted by $X^\mu$ on which the fields may depend. $a,b=0,\cdots, 9$
are local Lorentz frame indices. We use the symbols $G, B, Q, S,
R$ to denote $10\times 10$ matrices and $\cal{G}, \cal{B}, \cal{Q},
\cal{S}, \cal{R}$ to denote their $d\times d$ blocks labeled by $i,j$.

\noindent\underline{{\it Parameterization of $SO(d,d)$} :} It is well
known that the low-energy effective action for the massless NS-NS
fields, which are restricted to be independent of $X^i$, is invariant
under an $SO(d,d)$ group of transformations \cite{MV,SEN,HS,MS}. Not
all elements of this group have a non-trivial action on the background
fields \cite{SEN}: In fact, out of its $d(2d-1)$ elements, a
$d^2$-dimensional subgroup corresponds to general linear coordinate
transformations $GL(d)$ and another $d(d-1)/2$-dimensional subgroup
corresponds to constant shifts in $B_{ij}$, both of which are manifest
symmetries of the low-energy effective action even without restricting
the fields to be independent of $X^i$. To describe the action of these
elements on the fields, we do not really need the $O(d,d)$ formalism.
Only a $d(d-1)/2$-dimensional subgroup $SO(d-1,1)\times
SO(d-1,1)/SO(d-1,1)$ acts non-trivially on the background fields and
therefore, in this paper, we focus on the transformations generated by
this subgroup. Let ${\cal S, R} \in SO(d-1,1)$. We choose a basis in
which elements of $SO(d-1,1)\times SO(d-1,1)\subset SO(d,d)$ take the
form 
\beq 
O=\left(\ba{cc} S & 0 \\ 0 & R \ea\right) 
\equiv \left(\ba{cccc} {\cal S} & 0 & 0 & 0 \\ 
                       0 &{\bf 1}_{10-d} & 0 & 0 \\ 
                       0 & 0 & {\cal R} & 0 \\ 
                       0 & 0 & 0 &{\bf 1}_{10-d} \ea\right)\,\,,\qquad 
O^T \left(\ba{cc} \hat\eta & 0 \\ 0 & -\hat\eta \ea\right) O =
\left(\ba{cc} \hat\eta & 0 \\ 0 & -\hat\eta \ea\right)\,.
\label{SR}
\eeq
Note that the transformations are embedded in $10\times 10$ matrices
and we use a {\it hat} to distinguish between the $SO(d-1,1)$
invariant metric $\hat\eta_{MN}$ appearing above, and the Minkowski
metric $\eta_{ab}$ of the local Lorentz frame. The case ${\cal
S}={\cal R}$ corresponds to a coordinate rotation which is already
included in $GL(d)$. The non-trivial transformations are therefore
parameterized by $SO(d-1,1)\times SO(d-1,1)/SO(d-1,1)$ which is
obtained by restricting $O$ such that its only independent parameters
are contained in ${\cal S}^{-1}{\cal R}$ (for example, by setting
${\cal S}= {\bf 1}_d$). In the following we may simply write
$SO(d,d)$, understanding that we are only interested in its
non-trivial elements.

\noindent\underline{{\it Action on NS-NS backgrounds} :} The action of 
the $SO(d,d)$ group on the metric $G_{MN}$ and the antisymmetric
tensor field $B_{MN}$ is most conveniently written in terms of a
matrix $M$ which, in our conventions for the parameterization of the
group elements, is given by \cite{MV,SEN,HS,MS} 
\beq
M = \left(\ba{rr}
(\hat\eta^{-1}K+{\bf 1}_{10})\,G^{-1}\,(\hat\eta^{-1}K^T+{\bf 1}_{10}) &
-(\hat\eta^{-1}K+{\bf 1}_{10})\,G^{-1}\,(\hat\eta^{-1}K^T-{\bf 1}_{10}) 
\\[.3cm] 
-(\hat\eta^{-1}K-{\bf 1}_{10})\,G^{-1}\,(\hat\eta^{-1}K^T+{\bf 1}_{10}) &
(\hat\eta^{-1}K-{\bf 1}_{10})\,G^{-1}\,(\hat\eta^{-1}K^T-{\bf 1}_{10}) 
\ea\right)\,.
\label{M}
\eeq 
Here, $K=G + B$ and the sign of $B_{MN}$ is chosen such that the
worldsheet action has the standard form $\int d^2\sigma
K_{MN}\del_+X^M \del_-X^N$. The NS-NS fields $G, B$ and $\phi$ then
transform according to
\beq
\t M = O\, M\, O^T\,\,,\qquad 
e^{2\t \phi}=e^{2 \phi}\sqrt{\det\t G/\det G}\,.
\label{NS-NS}
\eeq
Using (\ref{M}) in (\ref{NS-NS}), one can check that the 
transformation of the metric $G$ can be written in two equivalent
forms \cite{SFH2},
\beq
\t G^{-1}=Q_-\,G^{-1}\,Q_-^T = Q_+\,G^{-1}\,Q_+^T \,\,, 
\label{G}
\eeq
where the matrices $Q^M_{-N}$ and $Q^M_{+N}$ are given by
\beq
\ba{cc}
Q_- =\frac{1}{2}\left[(S+R)+(S-R)\hat\eta^{-1}(G+B)\right]\,,
\\[.3cm]
Q_+ =\frac{1}{2}\left[(S+R)-(S-R)\hat\eta^{-1}(G-B)\right]\,.
\ea
\label{Qpm}
\eeq
These matrices will play an important role in our discussions later.  
Their relevance can be understood better by noting that the canonical
transformation that implements the $SO(d,d)$ transformation at the
worldsheet level has the form $\del_\pm \t X^M = Q^M_{\pm N}\del_\pm
X^N$, where we have ignored the worldsheet fermion contributions. 
On the worldsheet fermions, the action takes the form $\t\psi^M_\pm
=Q^M_{\pm N}\psi^N$ \cite{SFH3}. It is also useful to write the
inverses of $Q_{\pm}$ in terms of the variables of the transformed
theory as,  
\beq
\ba{cc} 
Q^{-1}_- =\frac{1}{2}\left[(S^{-1}+R^{-1})+(S^{-1}-R^{-1})
\hat\eta^{-1}(\t G+\t B)\right]\,,
\\[.3cm]
Q^{-1}_+ =\frac{1}{2}\left[(S^{-1}+R^{-1})-(S^{-1}-R^{-1})
\hat\eta^{-1}(\t G-\t B)\right]\,.
\ea
\label{Qinpm}
\eeq
Then, from the form of $S$ and $R$ given in (\ref{SR}) it is clear
that  
\beq 
(Q_\pm)^\mu_{\,j}=(Q^{-1}_\pm)^\mu_{\,j}=0
\,,\qquad (Q_\pm)^\mu_{\,\nu}=(Q^{-1}_\pm)^\mu_{\,\nu}=
\delta^\mu_{\,\nu}\,.
\label{Qcompo}
\eeq
In terms of the matrix $Q_-$ the dilaton transformation takes the from
\beq
e^{\phi-\t\phi} = \sqrt{\,\det{Q_-}}\,.
\label{phiQ}
\eeq

\noindent\underline{{\it Induced local Lorentz Twist} :} In order to
obtain the transformations of spinors and R-R fields, we also need the
transformation of the vielbein $e^a_{\,M}$. From now on, for
convenience, we will use the symbol $e$ for the inverse vielbein
$e^M_{\,a}$ and will refer to both $e$ and $e^{-1}$ as the vielbein.
From equation (\ref{G}) it is evident that {\it a priori} one can
transform the vielbein in two different ways \cite{SFH3},
\beq
\t e^M_{(-)a} = Q^M_{-N}\,e^N_{\,\,a}\,,\qquad
\t e^M_{(+)a} = Q^M_{+N}\,e^N_{\,\,a}\,, 
\label{epm}
\eeq
both leading to the same transformed metric $\t G^{-1} =
\t e \eta^{-1}\t e^{\,T}$. These two vielbeins are related by a local
Lorentz transformation $\L$,
\beq
\t e^M_{(+)b}=\t e^M_{(-)a} \, \L^a_{\,b}\,\,,\qquad
\L= e^{-1}Q^{-1}_-Q_+e\,.
\label{llt}
\eeq
The appearance of the two vielbeins $\t e_{(\pm)}$ can be easily
understood in terms of the worldsheet theory \cite{SFH4}: One may
regard the vielbein $e^M_a$ as originating in either the left-moving
or the right-moving sector of the worldsheet theory. It then
transforms to either $\t e_{(+)}$ or $\t e_{(-)}$, depending on its
worldsheet origin. This is consistent with the action of $SO(d,d)$ on
the worldsheet variables as described above and also with the fact
that the worldsheet parity $\sigma\rightarrow -\sigma$, which
interchanges the two worldsheet sectors, also interchanges the
variables $(S, R, B)$ and $(R, S, -B)$, and hence $\t e_{(+)}$ and $\t
e_{(-)}$.

It is then clear that an $SO(d-1,1)\times SO(d-1,1)/SO(d-1,1)$
transformation twists the local Lorentz frame originating in the
left-moving worldsheet sector with respect to the one originating in
the right-moving sector by an amount $\L^a_{\,b}$. Although all NS-NS
sector fields and their supersymmetry variations are insensitive to
$\L$, in general this twist can be absorbed by the Ramond sector
fields, leaving the two local Lorentz frames untwisted. In fact, as we
will see in section 4, this is dictated by the requirement of
consistency with supersymmetry which determines the way the spinor
representation of the induced twist $\L^a_{\,b}$ enters in the
transformations of space-time spinors and Ramond-Ramond fields under
non-trivial $SO(d,d)$ action. However, before that, we first construct
the spinor representation of the induced twist $\L$ in the next
section. 

\section{Spinor Representation of the Induced Twist}
Let us consider the $\G$-matrices $\G^M=e^M_{\,\,a}\G^a$ that are
associated with the local Lorentz frames that originate in both the
right- and the left-moving sectors of the worldsheet theory. After a
non-trivial $SO(d,d)$ transformation, these transform to either
$\t\G^M_{(-)}=\t e^M_{(-)a}\G^a$ or $\t\G^M_{(+)}=\t e^M_{(+)a}\G^a$,
depending on the worldsheet sector in which the Dirac algebra
originates. Let us choose $\t e^M_{(-)a}$ as the vielbein in terms of
which the transformed theory is to be finally written\footnote{Of
course, this is a matter of choice and we could as well choose $\t
e^M_{(+)a}$ or any other Lorentz equivalent vielbein. However, with
$\t e^M_{(-)a}$ as the vielbein, it is natural to set $S=1$ so that in
the flat-space limit, $\t e_{(-)}\rightarrow e$ while $\t e_{(+)} 
\rightarrow Re$.}and express $\t e^M_{(+)a}$ in terms of it, using
(\ref{llt}). Then we get
\beq 
\t\G^M_{(+)} = \O^{-1}\,\t\G^M_{(-)}\, \O\,,
\label{GpGm}
\eeq
where $\O$ is the spinor representation of the $SO(d,d)$ induced twist
$\L$ defined by
\beq
\O^{-1}\,\G^a\,\O = \L^a_{\,\,b}\,\G^b\,.
\label{srep}
\eeq
The factor $\O$ in (\ref{GpGm}) can now be absorbed in the
transformation of space-time spinors. Note that this only affects
spinor indices that originate in the left-moving Ramond sector of the
worldsheet theory, leaving the ones associated with the right-moving
Ramond sector unchanged. We will see in the next section that this is
in fact dictated by supersymmetry. This situation is quite reminiscent
of the construction of Ramond sector boundary states in the presence
of a background worldvolume gauge field studied in \cite{CNLY}. Below,
we explicitly construct the spinor representation $\O$ associated with
$SO(d,d)$. For the simpler case of a single T-duality, for which $det\,
O=-1$, see \cite{SFH4}.   

In general, $\L^a_{\,\,b}$ is an element of the local Lorentz group
$SO(9,1)$ associated with the left-moving sector of the worldsheet
theory. Therefore, it can be parameterized by an antisymmetric matrix
$A_{ab}$ as
\beq
\L=(\eta +A)^{-1}(\eta -A)\,,\qquad
A=\eta\,({\bf 1}-\L)\,({\bf 1}+\L)^{-1}\,.
\label{AL}
\eeq
Then, the antisymmetry of $A$ implies that $\L^T\eta\L=\eta$. For the
time being we assume that ${\bf 1}+\L$ is non-singular so that the
above parameterization is well defined. We will later come back to the
singular case. The spinor representation can now be constructed in
terms of $A$ as  
$$
\O=\left[\det(\eta+A)\right]^{-1/2}\AE\,(-\frac{1}{2}A_{ab}\G^{ab})\,,
$$  
where the symbol $\AE$ stands for an exponential-like expansion with
the products of $\G$-matrices fully antisymmetrized,
$$
\AE(-\frac{1}{2}A_{ab}\G^{ab})=
1 + \sum_{n=1}^{n=5}\frac{(-1)^n}{n!\,2^n}A_{a_1b_1}\cdots A_{a_nb_n}
\G^{a_1b_1\cdots a_nb_n}\,.
$$
{\it A priori}, $\O$ contains products of up to 10 $\G$-matrices.
However, note that equation (\ref{srep}) implies $\O^{-1}\G^M\O=
(Q^{-1}_-Q_+)^M_{\,\,N}\G^N$. Furthermore, from (\ref{Qcompo}) one can
see that $(Q^{-1}_-Q_+)^\mu_{\,\,\nu}=\delta^\mu_\nu$. This shows
that $\O$ keeps $\G^\mu$ invariant and therefore should be
constructed in terms of the $d$ matrices $\G_i$ alone, which
anti-commute with $\G^\mu$. To make this feature explicit, let us
define $A^{MN}=e^{Ma} e^{Nb}A_{ab}$. Then, using $\L$ as given by
(\ref{llt}) in the expression for $A$ in (\ref{AL}), we find that the
only non-zero components of $A^{MN}$ are given by
\beq
{\cal A}^{ij}=\left[\hat\eta_d({\bf 1}_d-{\cal S}^{-1}{\cal R})^{-1}
({\bf 1}_d +{\cal S}^{-1}{\cal R}) + {\cal B}\right]^{-1}_{ij}\,.
\label{A1}
\eeq
Here ${\cal A}$, ${\cal S}$, ${\cal R}$ and $\hat\eta_d$ denote the
$d\times d$ blocks of $A$, $S$, $R$ and $\hat\eta$ labeled by
$i,j=0,\cdots d-1$, and the $(ij)$ indices on the right hand side are
raised by the matrix inversion. Note that the first term in the square
bracket is the inverse of the antisymmetric matrix that parameterizes
the orthogonal transformation ${\cal S}^{-1}{\cal R}$ through
equations similar to (\ref{AL}). Substituting for the determinant
prefactor as well, the spinor representation takes the form 
\beq
\O=2^{-\frac{d}{2}}
\sqrt{\frac{\det({\cal Q}_- + {\cal Q}_+)}{\det{\cal Q}_-}}\,
\AE 
(-\frac{1}{2}{\cal A}^{ij}\G_{ij})\,,
\label{S-rep}
\eeq
where now,
\beq
\AE(-\frac{1}{2}{\cal A}^{ij}\G_{ij})=
1 + \sum_{p=1}^{[d/2]}\frac{(-1)^p}{p!\,2^p}{\cal A}^{i_1i_2}\cdots 
{\cal A}^{i_{2p-1} i_{2p}} \G_{i_1 i_2\cdots i_{2p-1} i_{2p}}\,.
\label{AE}
\eeq
Here, $[d/2]$ stands for the integer part of $d/2$ so that, for
example, the summation contains only one term for $d=2,3$ (involving
$\G_{i_1i_2}$) and two terms for $d=4,5$ (involving $\G_{i_1i_2}$ and
$\G_{i_1i_2i_3i_4}$). In particular, it does not contain $\G_\mu$ as
should be the case. The factor $\det{\cal Q}_-$ is the same quantity
that appears in the transformation of $e^{2\phi}$ and $\sqrt{\,\det
G}$. This will be important in the derivation of the transformation
rules of the Ramond-Ramond fields as well as in showing the invariance
of the low-energy effective action. The remaining factor has the
explicit form  
\beq
\det({\cal Q}_-+{\cal Q}_+)=\det[({\bf 1}_d+{\cal S}^{-1}{\cal R}) 
+ ({\bf 1}_d-{\cal S}^{-1}{\cal R})\hat\eta_d\,{\cal B}]\,,
\label{dQpQm}
\eeq
where we have used the fact that $\det{\cal S}=1$. As will be shown
below, this factor is essential for getting the correct T-duality
limit in cases with ${\cal B}_{ij}=0$, which serves as an easy
check of the results. Note that both this factor and ${\cal A}^{ij}$
depend only on ${\cal B}$ and have no dependence on $G$ and other
components of $B$. Furthermore, note that, as expected, $\O$ depends
on the combination ${\cal S}^{-1}{\cal R}$ alone and equals identity
for ${\cal S}={\cal R}$, corresponding to an $SO(d)\subset GL(d)$
transformation on the bosonic backgrounds which does not affect the
spinor index. One can easily restrict these formulae to the case when
$X^i$ do not contain the time coordinate by replacing $\hat\eta_d$ by
${\bf 1}_d$. 

\noindent\underline{{\it Singular Limits and T-duality with
$B_{ij}=0$} :} When $\det({\bf 1}+\L)=0$, the parameterization in
(\ref{AL}) becomes singular, though in the spinor representation $\O$,
the singular terms cancel out and one is left with a well defined
expression. This case is important as it includes T-duality
transformations in flat-space (as well as in curved backgrounds as
long as ${\cal B}_{ij}=0$) and provides an easy non-trivial check for
the correctness of our results.

To see this more clearly, let us consider, as an example, the $d=4$
case in the flat background $G=\eta$, $B=0$, which remains
unchanged by the transformation. We assume all the four $X^i$ to be
spatial coordinates and set $S=1$. Then, $\L=R$. T-duality
transformations along the four coordinates $X^i$ correspond  
to taking ${\cal R}=-{\bf 1}_4$, where ${\cal R}$ is the $4\times 4$ 
block of $R$. In this case the spinor representation is already known 
(see, for example, \cite{POL}) and is simply given by
$\O=\G_1\G_2\G_3\G_4$. On the other hand, in the formalism here, this
transformation clearly corresponds to a singular case since
$\det(1+\L)=0$. One can easily see that the correct spinor
representation is reproduced by first going away from the T-duality
point and parameterizing ${\cal R}$ as
\beq
{\cal R}=
\left(\ba{cc}
\ba{cc} \cos\theta & \sin\theta \\ -\sin\theta & \cos\theta \ea   
& {\LARGE\rm O} \\
{\LARGE\rm O} & \ba{cc} \cos\phi & \sin\phi \\ -\sin\phi & \cos\phi \ea   
\ea\right)\,.
\label{R4}
\eeq
The singular cases then correspond to either $\theta=\pi$ or
$\phi=\pi$ or both. One can now easily evaluate $\sqrt{\det({\cal
Q}_+ + {\cal Q}_-)}$, which tends to zero in the singular limits, and
${\cal A}^{ij}$ (some components of which tend to $\infty$ in the
singular limits) and get a well defined $\O$ as  
$$
\O=\cos\frac{\theta}{2}\cos\frac{\phi}{2}+
\sin\frac{\theta}{2}\cos\frac{\phi}{2}\G_{12}+
\sin\frac{\phi}{2}\cos\frac{\theta}{2}\G_{34}+
\sin\frac{\theta}{2}\sin\frac{\phi}{2}\G_{1234}\,.
$$
By setting $\theta=\phi=\pi$ we reproduce the correct $\O$ for
T-dualities along all the four coordinates, while setting only one of
the angles to $\pi$ corresponds to T-dualities along two coordinates,
combined with a non-trivial $O(2,2)$ transformation involving the
other two. This discussion highlights the importance of the
$\sqrt{\det({\cal Q}_+ + {\cal Q}_-)}$ factor which is essential in
eliminating the terms in (\ref{AE}) - particularly the identity - that
should not appear in the flat-space T-duality limit.

Since neither $\sqrt{\det({\cal Q}_+ + {\cal Q}_-)}$ nor ${\cal
A}^{ij}$ depend on the metric, its inclusion does not change the
situation described above for the flat space, except that now $\O$
also contains the $(\det Q_-)^{-1/2}$ factor. For example, if we
consider two T-dualities along $X^1$ and $X^2$ with ${\cal B}_{12}=0$,
then $\O=(\det{\cal G})^{-1/2}\G_{12}$, as can be directly
verified using the result for a single T-duality in \cite{SFH4}.
This case is relevant to the application of $SO(d,d)$ transformations
to the simplest forms of Dp-brane solutions for which the NS-NS two
form vanishes. 

In the more general case when ${\cal B}\neq 0$, the singular limit
$\det({\cal Q}_+ + {\cal Q}_-)=0$ includes, in particular, the
interesting case ${\cal Q}_+=-{\cal Q}_-$, or ${\cal S}^{-1}{\cal R}=
-({\bf 1}_d+\hat\eta^{-1}_d{\cal B})({\bf 1}_d-\hat\eta^{-1}_d {\cal
B})^{-1}$, {\it i.e.}, when ${\cal B}$ parameterizes the orthogonal
matrix ${\cal R}^{-1}{\cal S}$ through equations similar to
(\ref{AL}). Clearly, this can arise only in even $d$ with constant
${\cal B}$ and the corresponding $\O$ will have only one term which is
proportional to $\G_{i_1\cdots i_d}$. Thus we see that in this case,
there exists a special $SO(d,d)$ transformation, the spinor
representation of which has the same form as that for $d$ T-dualities
in flat space. In general, even in non-flat backgrounds, one can
parameterize $\L^a_{\,\,b}$ as ${\cal R}$ in (\ref{R4}) to show that a
well defined spinor representation always exists in the singular
limits. Having constructed the spinor representation of the induced
twist, we set out to find the $SO(d,d)$ action on the massless NS-R
and R-R fields that involve $\O$.

\section{O(d,d) Transformation of space-time Spinors}

Let us consider the gravitinos $\Psi_{\pm M}$, the dilatinos
$\lambda_{\pm}$ and the space-time supersymmetry transformation
parameters $\e_\pm$ in type-II string theories. The subscripts
``$\pm$'' denote the worldsheet sector (``$+$'' for left-moving and
``$-$'' for right-moving) in which the spinor index of the space-time
spinor, or equivalently its Ramond component, originates. We assign
$positive$ chirality to $\e_-$ in both IIA and IIB theories which
fixes the chiralities of all other spinors. All spinors are
assumed to be independent of $d$ coordinates $X^i$ but may depend on
the remaining $10-d$ coordinates $X^\mu$. The action of the
non-trivial elements of the $O(d,d)$ group on these spinors can be
obtained by demanding consistency with supersymmetry transformations.
These supersymmetry transformations were constructed in \cite{Romans}
for type-IIA theory and in \cite{JHS} for type-IIB theory. Here, we
write the supersymmetry transformations in the string metric and in
conventions more natural to string theory as given in \cite{SFH4}.
Denoting the supersymmetry variations corresponding to $\e_+$ and
$\e_-$ by $\delta_+$ and $\delta_-$ respectively, it turns out that
$\delta_\pm \Psi_\pm$ and $\delta_\pm \lambda_\pm$ depend only on the
NS-NS fields, on which the $O(d,d)$ action is well known, and are
independent of R-R fields. This is expected from the fact that in
string theory the supercharges act independently on the left- and
right-moving worldsheet sectors, interchanging R and NS boundary
conditions. Therefore, we can use these variations to determine the
$O(d,d)$ action on the spinors. On the other hand, the variations
$\delta_\pm\Psi_\mp$ and $\delta_\pm \lambda_\mp$ depend on the R-R
fields alone and will be used to obtain their transformations in the
next section. In general, these results are valid for $O(d,d)$, and we
restrict ourselves to the $SO(d,d)$ subgroup only when using the
explicit construction of the spinor representation given in the
previous section. The derivation below closely follows the one in
section 3 of \cite{SFH4} and generalizes it from the single T-duality
case to the $O(d,d)$ group actions.

Defining the torsionful spin-connections as $W^\pm_{Mab}=w_{Mab}\mp
\frac{1}{2}H_{Mab}\,$, the gravitino supersymmetry variations 
$\delta_\pm \Psi_\pm$ are given by 
\bea
\delta_-\Psi_{- M}&=&\left(\del_M + \frac{1}{4}\,W^-_{Mab}\,
\G^{ab}\right)\e_-\,\,+\,\cdots \,,
\label{dG-}\\[.2cm]
\delta_+\Psi_{+ M}&=&\left(\del_M + \frac{1}{4}\,W^+_{Mab}\,
\G^{ab}\right)\e_+\,\,+\,\cdots \,,
\label{dG+}
\eea
where, ``$\cdots$'' denotes 3-spinor terms the explicit forms of which
are not needed here. To obtain the $O(d,d)$ action on $\Psi_{\pm M}$,
we need the transformations of $W^\pm_{Mab}$\,. In the $O(d,d)$
transformed theory, one can define two sets of torsionful spin-connections
: $\t W^\pm_{(-)Mab}$ corresponding to the vielbein $\t e_{(-)}$ and
$\t W^\pm_{(+)Mab}$ corresponding to the vielbein $\t e_{(+)}$, where
the two vielbeins are given by (\ref{epm}). One can then show that
(see appendix A)
\bea
\t W^-_{(-)Mab} &=& W^-_{Nab} (Q^{-1}_+)^N_{\,\,M}\,, 
\label{W--}\\[.2cm]
\t W^+_{(+)Mab} &=& W^+_{Nab} (Q^{-1}_-)^N_{\,\,M}\,.
\label{W++}
\eea 
Since we have chosen $\t e_{(-)}$ as the vielbein in terms of which
the transformed theory is to be written, the corresponding
supersymmetry variations should be written in terms of $\t
W^\pm_{(-)Mab}$ alone. Let us first consider $\delta_-\Psi_{- M}$.
Equation (\ref{Qcompo}), along with the fact that $\e_-$ is independent
of $X^i$, implies that $\del_M\e_-(Q^{-1}_+)^M_{\,\,N}=\del_N\e_-$.
Then multiplying (\ref{dG-}) by $(Q^{-1}_+)^M_{\,\,N}$ and using
(\ref{W--}) leads to the corresponding variation in the transformed
theory provided we identify $\t\e_-=\e_-$ and $\delta_-\t\Psi_{-
N}=\delta_-\Psi_{- M} (Q^{-1}_+)^M_{\,\,N}$, up to 3-spinor terms.
Similar steps applied to $\delta_+\Psi_{+ M}$, on the other hand, will
lead to 
\bea
\delta_+\Psi_{+ M}(Q^{-1}_-)^M_{\,\,N}&=&\left(\del_M + \frac{1}{4}\,
\t W^+_{(+)Mab}\,\G^{ab}\right)\e_+\,\,+\,\cdots \,,\nonumber\\[.2cm]
&=&\O^{-1}\left(\del_M + \frac{1}{4}\,\t W^+_{(-)Mab}\,\G^{ab}\right)
\O\,\e_+\,\,+\,\cdots \,.\nonumber
\eea
Here, we have used the fact that, due to (\ref{llt}), $\t W_{(+)M}$
is related to $\t W_{(-)M}$ by a local Lorentz transformation $\L\,$,
$\t W_{(+)M}=\L^{-1}\t W_{(-)M}\L + \L^{-1}\del_M\L$, which has a
spinor representation $\O$ defined by (\ref{srep}). The above equation
will lead to the correct form for the variation $\delta_+\t\Psi_{+ M}$
in the transformed theory provided we identify $\t\e_+=\O\,\e_+$ and 
$\delta_+\t\Psi_{+ N}=\O\,\delta_+\Psi_{+ M} (Q^{-1}_-)^M_{\,\,N}$,
again up to 3-spinor terms. Thus under an $O(d,d)$ action, the
supersymmetry transformation parameters transform to 
\beq
\t \e_-=\e_-\,\,,\qquad\qquad \t\e_+ = \O\,\e_+\,,
\label{Odd-epm}
\eeq
while the transformation of the gravitinos can be read off from the
$O(d,d)$ action on the supersymmetry variations $\delta_\pm\Psi_{\pm
M}$ as, 
\beq
\ba{rcl}
\t\Psi_{-M} & = &\Psi_{-N}\,(Q^{-1}_+)^N_{\,\,M}\,,\\[.3cm] 
\t\Psi_{+M} & = & \O\,\Psi_{+N}\,(Q^{-1}_-)^N_{\,\,M}\,. 
\ea
\label{Odd-G}
\eeq
Note that though the transformations we obtained for $\delta_\pm
\Psi_{\pm M}$ receive corrections cubic in the spinors, our final
results for $\Psi_{\pm M}$ do not receive such corrections and are, in
this sense, exact. This can be understood on general grounds by noting
that a 3-spinor correction to (\ref{Odd-G}) will induce 4- and higher
spinor couplings with derivative interactions in the supergravity
Lagrangian, which should not exist. More precisely, such corrections
are ruled out by the compatibility of the $O(d,d)$ transformations
with the supersymmetry variations of the NS-NS fields. This was shown
in \cite{SFH4} for the case of a single T-duality. The same argument
applies to the general case here simply because, in principle, any
non-trivial $O(d,d)$ transformation can be constructed by intertwining
T-dualities with $GL(d)$ transformations (an explicit construction,
though not needed for this argument, will be given later). Thus the
$O(d,d)$ transformations have the basic property that they do not mix
quantities with different space-time fermion numbers.   

Though the results we have so far are enough to derive the $O(d,d)$
action on the Ramond-Ramond fields, for the sake of completeness, we
also obtain the action on the dilatinos. The dilatino supersymmetry
variations $\delta_\pm\lambda_\pm$ are given by 
\beq
\delta_\pm \lambda_\pm= \frac{1}{2}\left(\G^M\del_M \phi \mp 
\frac{1}{12}\G^{MNK} H_{MNK}\right)\e_\pm\,\,+\cdots\,,
\label{dD-NS} 
\eeq 
where again ``$\cdots$'' denotes 3-spinor terms. To relate these
variations to the corresponding ones in the $O(d,d)$ transformed
theory, we need the transformation of $H_{MNK}$. It is most useful to
write this in the following two equivalent forms (see appendix A),
\beq
\t e^M_{(\pm) a}\t e^N_{(\pm) b}\t e^K_{(\pm) c} \t H_{MNK} =
e^M_a e^N_b e^K_c H_{MNK}  
- 3\left[Q^{-1}_\pm (S-R)\hat\eta^{-1}\right]^{ij}\,
W^\pm_{j[ab} e^{\hphantom\pm}_{c]i}\,.
\label{OddH}
\eeq
Using these and comparing $\delta_\pm\lambda_\pm$ with $\delta_\pm 
\t\lambda_\pm$ in the transformed theory, we get the $O(d,d)$ action
on the dilatinos as  
\beq
\ba{rcl}
\t\lambda_- &=& \lambda_- 
-\frac{1}{2} \G_i \left[ Q^{-1}_-(S-R)\hat\eta^{-1}\right]^{ij}
\Psi_{-j} 
\,, \\[.3cm] 
\t\lambda_+ &=& \O\left(\lambda_+
+\frac{1}{2} \G_i \left[ Q^{-1}_+(S-R)\hat\eta^{-1}\right]^{ij}
\Psi_{+j} \right)\,.
\ea
\label{OddD}
\eeq
Note that so far we have not restricted ourselves to supersymmetric
backgrounds for which the spinors and their supersymmetry variations
vanish. In such cases, the transformations above lead to the $O(d,d)$
action on the fermionic zero-modes on the background. 

\section{SO(d,d) Transformation of R-R Fields}
Having determined the $O(d,d)$ action on $\e_\pm$ and $\Psi_{\pm M}$,
we are now in a position to write down its action on the Ramond-Ramond
{\it field strengths} and {\it potentials}, again by demanding
consistency of the transformations with supersymmetry. For the
explicit construction of the transformations we restrict ourselves to
$SO(d,d)$. This will automatically insure the $SO(d,d)$ covariance of
the supergravity equations of motion as well as that of the
electric-magnetic duality constraints on the R-R fields since all
these are determined by supersymmetry. We will then compare our
results with the ones obtained in the framework of NCSYM approach to
Matrix theory compactifications for the $SO(3,3)$ action in type-IIA
on $T^3$. We will end the section by writing down an $SO(d,d)$
covariant form for the Ramond-Ramond kinetic terms in the low-energy
effective action. 

It is most convenient to start with the supersymmetry variation
$\delta_-\Psi_{+M}$ which involves the Ramond-Ramond field strengths.
In both IIA and IIB theories this can be written as \cite{Romans,JHS}
(see \cite{SFH4} for the required change of variables),
\beq
\delta_-\Psi_{+M}=\frac{1}{2(8)}\,e^\phi\, {\cal F}\,\G_M\,\e_-\,\,
+\,\cdots\,.
\label{dG-RR}
\eeq
Here, ``$\cdots$'' denote 3-spinor terms and ${\cal F}$ is a bispinor
that contains the Ramond-Ramond field strengths and has the expansion,
\beq
{\cal F}=\sum_n\frac{(-1)^n}{n!}\,F^{(n)}_{M_1\cdots M_n}\G^{M_1\cdots
M_n}\,.  
\label{bispin}
\eeq
The fact that $\e_-$ and $\Psi_{+M}$ have the same chirality in IIB
and opposite chiralities in IIA implies that in IIA theory the
summation contains only terms with even $n$ ($n=0,2,4,6,8,10$), while
in IIB theory it contains only terms with odd $n$ ($n=1,3,5,7,9$).
Note that, for the time being, we allow $n=0$ and $n=10$ in type IIA
theory, so that our formalism includes the massive type-IIA theory in
\cite{Romans}. The fact that $\e_-$ has definite chirality (in our
case $+1$ in both IIA and IIB) implies that ${\cal F}$ satisfies the
constraint ${\cal F}=-{\cal F}\,\G_{11}$ which, in terms of the
components fields, translates to $1/(10-n)!\,
F^{(10-n)}\,\G_{(10-n)}=1/n!\,F^{(n)}\,\G_{(n)}$. However, in the
following it is more convenient to retain both $F^{(n)}$ and
$F^{(10-n)}$ in the summation. Furthermore, we assume that $F_{M_1
\cdots M_n}$ are independent of the coordinates $X^i$, which is in
fact required by the $X^i$-independence of $\e_-$ and $\Psi_{+M}$ in
(\ref{dG-RR}). However, for the time being, we do not demand the
Ramond-Ramond potentials $C^{(n)}_{M_1\cdots M_n}$ to be $X^i$
independent as this would exclude the massive IIA theory (we will
briefly discuss this issue later). Let us now consider the
supersymmetry variation $\delta_-\t\Psi_{+M}$ in the theory obtained
after an $O(d,d)$ transformation,
\beq
\delta_-\t\Psi_{+M}=\frac{1}{2(8)}\,e^{\t\phi}\, \t{\cal F}\,\t\G_{(-)M}
\,\t\e_-\,\,+\,\cdots\,.
\label{dtG-RR}
\eeq
$\t{\cal F}$ has the same form as ${\cal F}$ above, with $F^{(n)}$
and $\G^{N}$ replaced by $\t F^{(n)}$ and $\t\G^{N}_{(-)}$,
respectively. Since $\t\G^N_{(-)}=Q^N_{-M}\G^M$, this becomes
\beq
\t{\cal F}=\sum_n\frac{(-1)^n}{n!}\,\t F^{(n)}_{N_1\cdots N_n}\,
Q^{N_1}_{-M_1}\cdots Q^{N_n}_{-M_n}\,\G^{M_1\cdots M_n}\,. 
\label{tbispin}
\eeq
Now, using (\ref{Odd-epm}) and (\ref{Odd-G}) along with (\ref{phiQ})
in (\ref{dtG-RR}) and comparing the result with (\ref{dG-RR}), we get
a compact expression for the $O(d,d)$ action on the Ramond-Ramond
bispinor field as 
\beq
\t{\cal F}= 
\sqrt{\,\det\,Q_-}\,\O\,{\cal F}\,.
\label{tF1}
\eeq
In other words, $e^\phi{\cal F}$, as well as $(\det G)^{1/4}{\cal F}$
transform as spinors under the induced Lorentz twist. This
expression is valid for $O(d,d)$ and could interchange IIA and IIB
theories for elements with determinant $-1$. However, to obtain the
transformation of the components of ${\cal F}$, we restrict ourselves
to the $SO(d,d)$ case and use the explicit expression for $\O$ as
given by (\ref{S-rep}). The above equation then reduces to
\beq
\t{\cal F}= 2^{-\frac{d}{2}}\sqrt{\det({\cal Q}_- + {\cal Q}_+)}\,
\,\AE\,(-\frac{1}{2}{\cal A}^{ij}\G_{ij})
\,{\cal F}\,.
\label{tF2}
\eeq
The expression for the component fields are obtained by fusing
together the antisymmetrized products of $\G$-matrices in the
expansions of $\AE$ (\ref{AE}) and ${\cal F}$ (\ref{bispin}),
using a useful $\G$-matrix identity in \cite{KHP,AVP} (that we quote 
in appendix B), and then matching the result with the expansion of 
$\t{\cal F}$ in (\ref{tbispin}). We will first consider a simple case
and then write down the general result.

\noindent\underline{{\it The $d=2,3$ Cases} :} Before writing down the
general result, it is instructive to look at the simplest case of
$[d/2]=1$ (corresponding to $SO(2,2)$ and $SO(3,3)$ transformations).
In this case, one can easily work out the transformation of
Ramond-Ramond field strengths, using the $\G$-matrix identity
(\ref{AppB2}) in appendix B, to obtain
\bea
\t F^{(n)}_{M_1\cdots M_n}&=&2^{-\frac{d}{2}}\sqrt{\det({\cal Q}_-+
{\cal Q}_+)}\, \Bigg[\,F^{(n)}_{N_1\cdots N_n}-\frac{1}{2}
{\cal A}^{i_1i_2}\Bigg( n(n-1)G_{i_1N_1}G_{i_2N_2}\,
F^{(n-2)}_{N_3\cdots N_n}
\nonumber\\[.2cm]
&&+\,2n G_{i_1N_1}\,F^{(n)}_{i_2N_2\cdots N_n}+
F^{(n+2)}_{i_2i_1N_1\cdots N_n}\,\Bigg)\,\Bigg]
(Q^{-1}_-)^{N_1}_{\,\,[M_1}\dots (Q^{-1}_-)^{N_n}_{\,\,M_n]}\,.
\label{O33F}
\eea
This expression is valid in both IIA and IIB theories, depending on
whether $n$ is $even$ or $odd$, and includes the massive IIA theory if
the field strengths $F^{(0)}$ or $F^{(10)}$ are non-vanishing. Also,
by construction, it does not mix $odd$-form and $even$-form field
strengths. However, it does transform the field strengths
non-trivially. In particular, if the original theory has only a
non-vanishing $q$-form field strength $F^{(q)}$, corresponding to
$D(q-2)$-branes, the theory obtained after the transformation will
generically have non-zero $q$, $(q-2)$ and $(q+2)$-from field
strengths, depending on whether the transformation involves directions
parallel or transverse to the brane, or a mixture of the two. On the
other hand, by construction, the new configuration preserves the same
amount of supersymmetry as the original solution. Thus the new
configuration can only be a real (non-marginal) bound state of $Dq$,
$D(q-2)$ and $D(q-4)$-branes. A more detailed discussion of this issue
can be found in \cite{BMM}.

As a simple explicit example, let us consider the case of non-trivial
$SO(2,2)$ transformations in flat-space. Setting $S=1$, this
corresponds to 
$$
{\cal R}=\left(\ba{cc} \cos\theta & \sin\theta \\
                  -\sin\theta & \cos\theta \ea\right)\,,\qquad 
{\cal A}^{ij}= \frac{\sin\frac{\theta}{2}}{\cos\frac{\theta}{2}}
\left(\ba{cc} 0 & -1 \\ 1 & 0 \ea\right)\,,
$$
with $Q_-=1$ and $\sqrt{\det({\cal Q}_-+{\cal Q}_+)}=2\cos(\theta/2)$.
Taking care of the antisymmetrization factors, one obtains
$$
\ba{ll}
\t F^{(n)}_{12\mu_3\cdots\mu_n}=\cos\frac{\theta}{2}\,F^{(n)}_{12\mu_3
\cdots\mu_n} +\, \sin\frac{\theta}{2}\,F^{(n-2)}_{\mu_3\cdots\mu_n}\,, 
\quad &
\t F^{(n)}_{1\mu_2\cdots\mu_n}=\cos\frac{\theta}{2}\,F^{(n)}_{1\mu_2
\cdots\mu_n} +\, \sin\frac{\theta}{2}\,F^{(n)}_{2\mu_2\cdots\mu_n}\,,
\\[.3cm]
\t F^{(n)}_{2\mu_2\cdots\mu_n}=\cos\frac{\theta}{2}\,F^{(n)}_{2\mu_2
\cdots\mu_n} -\, \sin\frac{\theta}{2}\,F^{(n)}_{1\mu_2\cdots\mu_n}\,,
\quad &
\t F^{(n)}_{\mu_1\cdots\mu_n}=\cos\frac{\theta}{2}F^{(n)}_{\mu_1
\cdots\mu_n} -\, \sin\frac{\theta}{2}\,F^{(n+2)}_{12\mu_1\cdots\mu_n}\,.
\ea
$$
In particular, for $\theta=\pi$ one reproduces the correct result for
two T-dualities along $X^1$ and $X^2$, as can be verified directly by
using the T-duality transformation of $F^{(n)}$ as given, for example,
in \cite{SFH4}.  Furthermore, the first equation above implies that if
$F^{(0)}$ is non-zero, then one gets a non-zero $\t F^{(2)}_{12}$. 
Therefore, in the case of massive IIA theory, one cannot restrict the
potential $C^{(1)}$ to be $X^i$ independent. This is discussed in more
detail in \cite{BRGPT} for a single T-duality and in \cite{CLPS} for
two T-dualities.  

\noindent\underline{{\it The General $SO(d,d)$ Case} :} 
Following the same steps as described above, one can work out
the transformation of Ramond-Ramond field strengths under the action
of non-trivial elements of the $SO(d,d)$ group, for generic $d\,$. Using
(\ref{AppB2}) in (\ref{tF2}), and after some straightforward
manipulations, one obtains   
\bea
&&\hspace{-1cm}
\t F^{(n)}_{M_1\cdots M_n}=
2^{-\frac{d}{2}}\sqrt{\det({\cal Q}_- + {\cal Q}_+)}\,
\sum_{p=0}^{[d/2]}\Bigg[ \frac{(-1)^p}{p! 2^p}\,
{\cal A}^{[i_1i_2}\cdots {\cal A}^{i_{2p-1}i_{2p}]}
\nonumber\\[.1cm]
&&\hspace{-.6cm}\times\left(\sum_{r=0}^{2p}\frac{(2p)!}{(2p-r)!}\,{}^nC_r\,
G_{i_1N_1}\cdots G_{i_rN_r}F^{(n+2p-2r)}_{i_{2p}\cdots i_{r+1}N_{r+1}
\cdots N_n}\right)\Bigg]\,
(Q^{-1}_-)^{N_1}_{\,\,[M_1}\cdots (Q^{-1}_-)^{N_n}_{\,\,M_n]}\,,
\label{SOdd-F}
\eea
where, ${}^nC_r$ are the binomial expansion coefficients and ${\cal
A}^{ij}$ and $\det({\cal Q}_- + {\cal Q}_+)$ are given by (\ref{A1})
and (\ref{dQpQm}), respectively. To evaluate $Q^{-1}_-$, one can
use either (\ref{Qpm}) or (\ref{Qinpm}), according to convenience.  
Besides the non-trivial elements that we have considered so far, the
full $SO(d,d)$ group also includes $GL(d)$ transformations involving
$X^i$ and constant shifts in $B_{ij}$. $F^{(n)}$ simply transform as
$n$-forms under the first set and are invariant under the second set. 
This completes the construction of the full $SO(d,d)$ action on the
Ramond-Ramond field strengths. 

Thought the transformation formula for $F^{(n)}$ looks rather bulky,
in practice, at least some components of it are simplified when
restricted to specific solutions. This is especially the case when the
potentials $C^{(n)}$ can be chosen to be $X^i$ independent, so that
$F_{i_1\cdots i_n}=0$.

\noindent\underline{{\it Transformation of Ramond-Ramond Potentials}
 :} 
So far, we have only demanded the $X^i$-independence of the R-R field
strengths and not that of the potentials since this would exclude the
massive type-IIA theory. This is pointed out above for the $SO(2,2)$
case in flat space and was considered in more detail in
\cite{BRGPT,CLPS}. Let us now restrict ourselves to the standard
type-IIA and IIB theories by setting the IIA mass parameter $F^{(0)}$
to zero. The $X^i$-independence of the Ramond-Ramond potentials is now
compatible with $SO(d,d)$ transformations.

It is convenient to define the potentials $C^{(n)}$ such that  
\beq
F^{(n)}_{M_1\cdots M_n} = n\del_{[M_1} C^{(n-1)}_{M_2\cdots M_n]}
-\frac{n!}{3!(n-3)!}H_{[M_1M_2M_3} C^{(n-3)}_{M_4\cdots M_n]}\,.
\label{RRpot}
\eeq
In this convention, the potentials are invariant under the NS-NS
$2$-form gauge transformations, but $C^{(4)}$ is not invariant under
the $SL(2,R)$ group of type-IIB theory. To obtain the $SO(d,d)$ action
on the Ramond-Ramond potentials, the straightforward (although
tedious) procedure would be to substitute (\ref{RRpot}) in
(\ref{SOdd-F}), and use the transformation of $H$ as given by
(\ref{OddH}), to work out the transformation of $C^{(n)}$ iteratively
in $n$. The calculation may get somewhat simplified if one chooses to
work in the local Lorentz frame, since then one will not have to
bother about the factors of $Q_-^{-1}$ in (\ref{SOdd-F}). However,
here we will obtain this transformation by a very simple argument: 

First, note that any non-trivial $SO(d,d)$ transformation can, in
principle, be constructed as a combination of appropriately chosen
discrete T-duality transformations and simple coordinate rotations
obtained by setting $S=R=A$. Since this statement is crucial for our
argument here and the one in the next section, we will spell it out in
some detail: Any $O(d)$ rotation, say $R$, can be decomposed as a
product of reflections about planes perpendicular to properly chosen
axes, or equivalently, as a product of reflections $T_i$ about planes
perpendicular to the coordinate axes $x^i$, and properly chosen
rotations $A_k$; $R=T_{i_n}A_{k_n}\cdots T_{i_1} A_{k_1}$. If we
choose $S=A_{k_n}\cdots A_{k_1}$, then the $O(d,d)$ transformation
implemented by $R$ and $S$ corresponds to a sequence of T-duality
transformations $T_{i}$ intertwined with coordinate rotations $A_k$,
which proves the statement above. Therefore, knowing how $F^{(n)}$
transforms under a single T-duality, one can in principle construct
its transformation under any non-trivial $SO(d,d)$ element by using
this decomposition, and reproduce equation (\ref{SOdd-F}). However, as
emphasized in \cite{SFH4}, $F^{(n)}$ and $C^{(n)}$ transform in
exactly the same way under a single T-duality (this is true up to a
sign which is immaterial since we need even number of T-dualities).
Being $n$-forms, they also transform in the same way under coordinate
transformations. This implies that if we construct the action of an
$SO(d,d)$ element on $C^{(n)}$, using its decomposition in terms of
T-dualities and rotations, then we will end up with the same
expression as for $F^{(n)}$. Hence,
\bea
&&\hspace{-1cm}
\t C^{(n)}_{M_1\cdots M_n}=
2^{-\frac{d}{2}}\sqrt{\det({\cal Q}_- + {\cal Q}_+)}\,
\sum_{p=0}^{[d/2]}\Bigg[ \frac{(-1)^p}{p! 2^p}\,
{\cal A}^{[i_1i_2}\cdots {\cal A}^{i_{2p-1}i_{2p}]}
\nonumber\\[.1cm]
&&\hspace{-.6cm}\times\left(\sum_{r=0}^{2p}\frac{(2p)!}{(2p-r)!}
\,{}^nC_r\,
G_{i_1N_1}\cdots G_{i_rN_r}C^{(n+2p-2r)}_{i_{2p}\cdots i_{r+1}N_{r+1}
\cdots N_n}\right)\Bigg]\,
(Q^{-1}_-)^{N_1}_{\,\,[M_1}\cdots (Q^{-1}_-)^{N_n}_{\,\,M_n]}\,.
\label{SOdd-C}
\eea 
As for the field strengths, $C^{(n)}$ are invariant under constant
shifts of $B_{ij}$ and transform as $n$-forms under $GL(d)$. This
completes the construction of the $SO(d,d)$ action on Ramond-Ramond
potentials. 
Defining a bispinor 
${\cal C}$ in the same way as ${\cal F}$ in (\ref{bispin}), the above
transformation takes the compact form
\beq
[\det \t G]^{\frac{1}{4}}\,\t{\cal C} =
\Omega\,[\det G]^{\frac{1}{4}}\,{\cal C} 
\label{Odd-C}
\eeq

We will not discuss the $SO(d,d)$ action on the R-R potentials in the
massive IIA theory, in which case some of the $C^{(n)}$ will
necessarily have an $X^i$ dependence and the above transformations
will get modified. However, we comment that, as shown in \cite{SFH4}
for the single T-duality case, it should be possible to define new
$X^i$ independent variables $\widehat C^{(n)}$ in terms of $C^{(n)}$ and
the mass parameter, such that the $SO(d,d)$ action on the potentials
has exactly the same form as in (\ref{SOdd-C}), but now with the
$C^{(n)}$ replaced by the new variables $\widehat C^{(n)}$.

\noindent\underline{{\it Comparison with Results from NCSYM} :} In some
cases, the transformation of the R-R potentials has also been studied
in the framework of M-theory compactifications to super Yang-Mills
theories on non-commutative tori \cite{BMZ,KS}, where the string
theory $SO(d,d)$ transformations are implemented as a Morita
equivalence. From the $10$-dimensional string theory point of view,
the transformation obtained in this approach is valid only in the
$G_{ij}\rightarrow 0$ limit and therefore, is not expected to coincide
with the ones given above. To make a comparison with these results, we
restrict ourselves to the case considered in \cite{BMZ}, which studies
the $1$-form and $3$-form potentials in type IIA theory compactified
on a 3-torus. The R-R potentials are assumed to have non-zero
components only along the torus directions. In our approach, the
$SO(3,3)$ transformation of these potentials can be read off from
(\ref{SOdd-C}) (or more directly from (\ref{O33F}) after replacing
$F^{(n)}$ by $C^{(n)}$) as,  
$$
\ba{l}
{\ds\frac{2^{3/2}\,\t C_i}{\sqrt{\,\det({\cal Q}_++{\cal Q}_-)}}}      
=\Bigg[C_l\,-\,\frac{1}{2}{\cal A}^{pq}\Bigg(2G_{pl}\,C_q -C_{pql}
\Bigg)\Bigg] (Q^{-1}_-)^l_{\,i}\,,\\[.5cm]
{\ds\frac{2^{3/2}\,\t C_{ijk}}{\sqrt{\,\det({\cal Q}_++{\cal Q}_-)}}} 
=\Bigg[C_{lmn}\,-\,{\cal A}^{pq}\Bigg(3\,G_{pl}\,G_{qm}\,C_n
+3\,G_{pl}\,C_{qmn}\Bigg)\Bigg](Q^{-1}_-)^l_{\,i} (Q^{-1}_-)^m_{\,j} 
(Q^{-1}_-)^n_{\,k}\,.
\ea
$$
The corresponding equations obtained in the NCSYM framework do not
coincide with these. However, in the limit $G_{ij}\rightarrow 0$, we
have 
$$
{\cal Q}_+={\cal Q}_-\equiv {\cal Q} \,,\qquad 
{\cal A}^{pq}=\frac{1}{2}[{\cal Q}^{-1}({\cal S}-{\cal R})]^{pq}\,,
\qquad ({\mbox as}\qquad G_{ij}\rightarrow 0)
$$
and the above transformations go over to, 
$$
\ba{l}
\t C_i \rightarrow \sqrt{\,\det(Q)}\left[\,C_l\,(Q^{-1})^l_{\;i}+
\frac{1}{4}\,C_{lmn}\,(Q^{-1})^l_{\;i}(Q^{-1})^m_{\;j}(Q^{-1})^n_{\;k}
\left[({\cal S}-{\cal R})\,Q^T\right]^{jk}\right] \,,\\[.4cm]
\t C_{ijk} \rightarrow \sqrt{\,\det(Q)}\,\, C_{lmn}\,
(Q^{-1})^l_{\;i}(Q^{-1})^m_{\;j}(Q^{-1})^n_{\;k}\,.
\ea
$$
These coincide with the transformations obtained in the NCSYM approach
\cite{BMZ,KS} (also see \cite{PS}). We can easily generalize the results
obtained so far in the NCSYM approach by taking the $G_{ij}\rightarrow
0$ limit in (\ref{SOdd-C}).  

\noindent\underline{{\it $SO(d,d)$ Covariant form of the R-R Kinetic
Energy Terms} :} Our analysis so far has been based on showing the
$SO(d,d)$ covariance of the supersymmetry transformations in type-II
theories. Since the equations of motion in these theories are
determined by supersymmetry, this approach also guarantees their
$SO(d,d)$ covariance, and hence that of the corresponding low-energy
effective actions. The $SO(d,d)$ covariant form of the effective
action for the NS-NS fields has been known for a long time
\cite{MV,SEN,HS,MS}. Here, we write down an $SO(d,d)$ covariant form
for the kinetic energy terms of the Ramond-Ramond fields in the
type-II low-energy effective actions (ignoring the subtlety associated
with the self-dual $5$-form in type-IIB theory).

In terms of the Ramond-Ramond bispinor ${\cal F}$ given by
(\ref{bispin}), The R-R kinetic energy terms in both IIA and IIB
theory can be written as 
\beq
\frac{1}{8}\sqrt{\det(G)}\;\sum_n \frac{1}{n!}\,F_{a_1\cdots a_n}\, 
F^{a_1\cdots a_n} = -\frac{2^{-5}}{8}\sqrt{\det(G)}\;
{\rm Tr}\left(\G^0\,{\cal F}^T\,\G^0\,\G_{11}\,{\cal F}\,\G_{11}
\right)\,, 
\label{RR-KE}
\eeq 
which can be checked using the trace formula (\ref{AppB3}) for the
$\G$-matrices given in appendix B. The index on $\G^0$ is a local
Lorentz frame index. The right hand side is simply a generalization of
the expression $\bar\psi\G_{11}\psi$ for spinor $\psi$ to the bispinor
${\cal F}$. In type-IIA theory $\G_{11}\,{\cal F}\,\G_{11}={\cal F}$
while in IIB $\G_{11}\,{\cal F}\,\G_{11}=-{\cal F}$, which restricts
the summation on the right hand side of the equation to even $n$ or
odd $n$, respectively. The equation includes $F^{(n)}$ and
$F^{(10-n)}$ separately and the duality between the two, including the
self-duality of the $5$-form, should be imposed by hand. The $\G_{11}$
factors in the expression on the right hand side have been inserted so
that the kinetic energy terms for $odd$-forms and $even$-forms have
the same sign, as should be the case in our metric convention which is
$\{-,+,\cdots,+\}$. Now, using (\ref{tbispin}) and (\ref{G}), one can
easily check that the R-R kinetic energy terms, as given by the left
hand side of (\ref{RR-KE}), are manifestly $SO(d,d)$ covariant since,
$$
\O^{-1} = -\G^0\,\O^T\,\G^0\,,\qquad \O\,\G_{11}=\G_{11}\,\O\,.
$$    
The Ramond-Ramond fields also enter the action through Chern-Simons
terms that we do not consider here, though it should be possible to
write these too in a manifestly $SO(d,d)$ covariant form using the
bispinor ${\cal F}$ and a similar quantity constructed in terms of the
R-R potentials. 

The $SO(d,d)$ covariant form of the R-R kinetic terms given above is,
as such, not restrictive enough to determine the $SO(d,d)$
transformation of the R-R field strengths if we did not already know
the transformation. However, the addition of the Chern-Simons terms may
change the situation. Another option is to use the kinetic terms
alone, but express the field strengths in terms of the potentials and
the NS-NS $3$-form field strength $H$. This form, along with the 
transformation of $H$ given in (\ref{OddH}) is restrictive enough to
determine transformation of the R-R potentials. In particular, note
that the antisymmetric part of the factor $Q^{-1}_\pm(S-R)
\hat\eta^{-1}$ appearing in (\ref{OddH}) is related to ${\cal A}^{ij}$
appearing in the transformation of $C^{(n)}$ in (\ref{SOdd-C}).

The above $SO(d,d)$ invariant form of the action is not invariant
under $O(d,d)$ transformations with $\det O=-1$ (including single
T-dualities) that interchange IIA and IIB theories. This is because in
such cases, $\O\,\G_{11}=-\G_{11}\,\O$ and after the transformation
one obtains the kinetic energy terms with the wrong sign.

\section{R-R Potentials as SO(d,d) Spinors}
We have seen that $O(d,d)$ transformations introduce a twist $\Lambda$
between the Lorentz frames associated with the left- and the
right-moving sectors of the worldsheet theory. This twist could be
regarded as an element of the space-time Lorentz group $O(9,1)$,
though its action differs from that of the Lorentz group in that it
affects only part of the fields. It was shown that the spinor
representation $\Omega$ of this twist determines the $O(d,d)$
transformations of R-R field strengths and potentials in a natural
way. Though the approach we have followed appears very natural from
the perturbative string theory point of view, there also exists, as
described in the introduction, an alternative approach to the problem
in which the R-R fields are arranged as components of an $SO(d,d)$
spinor \cite{HT,EW,AAFFT,BMZ,FOT,KS}. While this construction may
appear {\it ad hoc} from the point of view of perturbative string
theory, it fits naturally in the U-duality group structure of
non-perturbative string theory. However, so far, it has not been
possible to obtain the R-R field transformations (\ref{SOdd-F}) and
(\ref{SOdd-C}) in this approach which, as we shall see, may not be an
easy task. In the following, we will discuss the relationship between
these two approaches. 

Let ${\bf F}$ and ${\bf C}$ denote the sums of n-forms
$\sum_{n=1}^{n=9} F^{(n)}$ and $\sum_{n=0}^{n=8} C^{(n)}$,
respectively. Equation (\ref{RRpot}) can then be written as ${\bf F}=
d {\bf C} - {\bf H}\wedge {\bf C}$ and the potentials $C^{(n)}$ defined
by it are invariant under the $B_{MN}$ gauge transformations. It is
also common to use an alternative set of potentials $C\,'^{(n)}$ defined
by  
\beq
{\bf C}\,'\,=\,{\bf C}\wedge e^{-{\bf B}}\,,\qquad 
{\bf F}= e^{\bf B}\wedge d {\bf C}\,'\,.
\label{Cprime}
\eeq
These are not invariant under the $B_{MN}$ gauge transformations.
Comparing with the construction in \cite{BMZ,FOT}, it is clear that
${\bf C}\,'$ is the field in terms of which the $SO(d,d)$ spinor is
constructed. We will briefly review this construction here: Let
$\gamma^I$, for $I=1,\cdots, 2d$, denote Gamma matrices of the Dirac
algebra associated with $SO(d,d)$ in a basis in which the invariant
metric is of the from ${\ds diag(\hat\eta_d,-\hat\eta_d)}$.
Defining ${\ds a^{i\pm}=(\gamma^i\pm\gamma^{d+i})/2}$, for
$i=1,\cdots,d$, one can easily verify that $a^{i+}$ and $a_j^{-}= 
\hat\eta_{jk}a^{k-}$ satisfy the Heisenberg algebra ${\ds \{a^{i+},
a_j^{-}\}=\delta^i_j}$. A basis for the spinor representation (Majorana
in this case) is then obtained by applying the {\it raising} operators
$a^{i+}$ on the {\it vacuum} state $|0\rangle$, defined by
$a_i^-|0\rangle =0$, for all $i$. Restricting to states with either
even or odd numbers of $a^+$'s, one obtains the two Majorana-Weyl
basis states. Now consider the components ${\ds C\,'^{(n)}_{\mu_1
\cdots\mu_q i_1\cdots i_{n-q}}}$ of R-R potentials, for all $n$ and 
fixed $q$. The $SO(d,d)$ Majorana-Weyl spinors constructed out of the
R-R potentials are then given by 
\beq
|{\bf C}\,'\rangle_{\mu_1\cdots\mu_q}=
\sum_{p=0}^{d} C\,'^{(q+p)}_{\mu_1\cdots\mu_q i_1\cdots i_{p}}
a^{i_1+}\cdots a^{i_p+}|0\rangle
\label{RRspinor}
\eeq
Note that since $n=q+p$ is either even (IIB) or odd (IIA), in the
summation on the right hand side $p$ runs over either even or odd
values at a time, ensuring that the spinor is Majorana-Weyl. The
$SO(d,d)$ action on $C\,'^{(n)}$ can be obtained by constructing the 
spinor representation of the group following the procedure described
in section 3. However, to elucidate the relationship to the approach
followed in this paper, we obtain the transformation of $C\,'^{(n)}$
under non-trivial $SO(d,d)$ elements in a much simpler way, using our
results in section 5.

The straightforward procedure of obtaining the transformations of
$C\,'^{(n)}$ from that of $C^{(n)}$, using (\ref{Cprime}), runs into
trouble since the field $B_{MN}$ transforms in a rather complicated
way \cite{SEN}. However, we can circumvent this and obtain these
transformations by a simple argument: Under single T-duality
transformations both $C^{(n)}$ and $B_{MN}$ transform in a much
simpler way. Using these transformations (for example, as given in  
\cite{SFH4}) in (\ref{Cprime}) one can easily work out the
transformation of $C\,'^{(n)}$ under a single T-duality, say along
$X^9$, to obtain
\beq
\t C\,'^{(n)}_{9i_2\cdots i_n}= C\,'^{(n-1)}_{i_2\cdots i_n}
\,,\qquad
\t C\,'^{(n)}_{i_1i_2\cdots i_n}= C\,'^{(n+1)}_{9i_1\cdots i_n}\,.
\eeq
Note that this is exactly how $C^{(n)}$ would transform in flat space.
Now, using the construction described above equation (\ref{SOdd-C}), we
can intertwine single T-dualities with coordinate rotations to obtain
non-trivial $SO(d,d)$ transformations of $C\,'^{(n)}$. On the other
hand, since $C\,'^{(n)}$ and $C^{(n)}$ transform in the same way under
coordinate rotations, this $SO(d,d)$ transformation of $C\,'^{(n)}$
can also be obtained from that of $C^{(n)}$ (\ref{SOdd-C}) in the
flat-space limit of $G_{MN} \rightarrow \hat\eta_{MN}$ and $B_{MN}=0$
\footnote{The {\it hat} on $\eta$ reminds us that the transformation
for $C\,'^{(n)}$ obtained this way is still in curved space and that
space-time indices are raised and lowered with $G_{MN}$.}. After some
rearrangements, and setting ${\cal S}={\bf 1}_d$ by a coordinate
rotation, we obtain,  
\bea 
&&\hspace{-1cm}
\t C\,'^{(n)}_{M_1\cdots M_n}=
2^{-\frac{d}{2}}\sqrt{\det({\bf 1}_d + {\cal R})}\,
\sum_{p=0}^{[d/2]}\Bigg[ \frac{(-1)^p}{p! 2^p}\,
\Theta_{[i_1i_2}\cdots \Theta_{i_{2p-1}i_{2p}]}
\nonumber\\
&&\hspace{-.6cm}\times\left(\sum_{r=0}^{2p}\frac{(2p)!}{(2p-r)!}\,{}^nC_r\,
\hat\eta^{i_{2p}j_{2p}}\cdots\hat\eta^{i_{r+1}j_{r+1}}\,
C\,'^{(n+2p-2r)}_{j_{2p}\cdots j_{r+1}[M_{r+1}\cdots M_n}\,
\delta^{i_1}_{\,\,M_1}\cdots\delta^{i_r}_{\,\,M_r]}\right)\Bigg]\,,
\label{SOdd-C'}
\eea 
with 
\beq
\Theta_{ij}=\left[
\hat\eta_d\,({\bf 1}_d +{\cal R})^{-1}({\bf 1}_d-{\cal R})
\right]_{ij}\,.
\label{Aflat}
\eeq
Clearly, $\Theta$ is the antisymmetric matrix that parameterizes
non-trivial $SO(d,d)$ elements in (\ref{SR}) (with ${\cal S}={\bf
1}_d$) and in terms of which the spinor representation of the group
can be constructed as described in section 3. The dependence on ${\cal
S}$ can be easily restored by performing a coordinate rotation by an
amount ${\cal S}^{-1}$ and at the same time replacing ${\cal R}$ with 
${\cal S}^{-1}{\cal R}$, as should be the case. 

We have obtained equation (\ref{SOdd-C'}) without having to assume
that the R-R potentials $C\,'^{(n)}$ transform as $SO(d,d)$ spinors. 
However, the final result can also be written in terms of the spinors
(\ref{RRspinor}) as  
\beq
|\t{\bf C}\,'\rangle_{\mu_1\cdots\mu_q}=\Omega_{NT}
|{\bf C}\,'\rangle_{\mu_1\cdots\mu_q}\,,
\eeq
where $\Omega_{NT}$ stands for the spinor representation of the
non-trivial elements of the $SO(d,d)$ parameterized by
$SO(d-1,1)\times SO(d-1,1)/SO(d-1,1)$. Thus, the explicit
transformation of $C^{(n)}$ (\ref{SOdd-C}) leads us, by a very simple
argument, to the transformation of $C\,'^{(n)}$ as $SO(d,d)$ spinors
(at least for the non-trivial part of the group). The converse,
however, is not true. To construct the transformation of $C^{(n)}$
from that of $C\,'^{(n)}$, we have to use (\ref{Cprime}) but the
$SO(d,d)$ transformation of $B_{MN}$ \cite{SEN} is expected to make
things complicated.

As we have seen in the previous section, $C^{(n)}$ and $F^{(n)}$
transform in exactly the same way under $SO(d,d)$ transformations. 
This implies that we can also construct $SO(d,d)$ spinors out of
$F^{(n)}$. In fact, in analogy with ${\bf C}\,'$ in (\ref{Cprime}), we
can define
\beq
{\bf F}\,'\,=\,{\bf F}\wedge e^{-{\bf B}}\,=\,d {\bf C}\,'\,.
\label{Fprime}
\eeq
Then $|{\bf F}\,'\rangle_{\mu_1\cdots\mu_q}$ constructed in analogy
with (\ref{RRspinor}) also transform as $SO(d,d)$ spinors. 

\section{Conclusions}

We have obtained the $SO(d,d)$ transformations of the Ramond-Ramond
field strengths and potentials and, in the process, have also
determined the transformations of the space-time supersymmetry
parameters, the gravitinos and the dilatinos in type-II theories. The
derivation is based on supersymmetry and is therefore guaranteed to be
consistent with the equations of motion provided the fields are
independent of the $d$ coordinates with respect to which the
transformation is performed. The transformations we obtain for the R-R
field strengths also include the massive IIA theory, though for the
R-R potentials we restrict ourselves to the usual ``massless'' IIA
case. Besides the general cases, we also discuss some special cases to
highlight some features of the transformation and to check the
correctness of our results in these cases. Since the transformations
could also include the time direction, we have been careful to keep
track of the indices by explicitly retaining the flat metric
$\hat\eta$. Though we restrict ourselves to the $SO(d,d)$ case, all
formulas which do not involve the explicit form of the spinor
representation $\O$, constructed in section 3, are also valid for the
$O(d,d)$ case. In the $SO(3,3)$ case we reproduce the results obtained
in the NCSYM formalism of Matrix theory compactifications by taking
the limit $G_{ij}\rightarrow 0$. It is also shown that the R-R kinetic
energy terms in the low-energy effective action can be easily written
in an $SO(d,d)$ invariant form in terms of the R-R bispinor. We also
clarify the relation between our approach and an alternative one which
is based on constructing Majorana-Weyl spinor representations of the
$SO(d,d)$ group in terms of the Ramond-Ramond potentials.   

The picture emerging is that $O(d,d)$ transformations induce a
rotation between the local Lorentz frames originating in the left- and
the right-moving sectors of the worldsheet theory. The misalignment
of the frames can be absorbed by the spinor index associated with one
of the frames. This is the basic mechanism by which the $O(d,d)$
action is transfered to the Ramond sector of the Theory.  

The transformations obtained in this paper can be used to construct
non-trivial D-brane configurations, starting from simpler ones. An
important feature of the transformation is that, applied to brane
configurations for which the surviving supersymmetries are independent
of $X^i$, it can produce more complicated brane configurations without
reducing the amount of supersymmetry preserved. This is an indication
that the final configuration corresponds to a non-trivial bound state
of branes, since otherwise it would invariably break some of the
supersymmetries. Though we have not considered such applications here,
as an example, we mention \cite{DMWY}. Here, the authors construct a
class of $SO(4,4)$ transformations by explicitly intertwining
T-dualities and spatial rotations to obtain a (D1,D5)-brane system
with a non-vanishing B-field. This solution corresponds to a genuine
bound state of D1 and D5-branes as opposed to the marginal bound state
with zero B-field. Some other examples were studied earlier in
\cite{BMM}

\vspace{-.2cm} 
\section*{Acknowledgments}
I would like to thank C. Angelantonj, I. Antoniadis and A. Armoni   
for useful discussions and comments. 

\vspace{-.2cm} 
\newcommand{\resection}[1]{\setcounter{equation}{0}\section{#1}}
\newcommand{\appsection}[1]{\setcounter{equation}{0}\section*{Appendix}}
\renewcommand{\theequation}{\thesection.\arabic{equation}}
\appendix
\appsection

\resection{Some derivations}
In this appendix we briefly describe the steps one could follow to
derive equations (\ref{W--}), (\ref{W++}) and (\ref{OddH}). To derive
(\ref{W--}), one starts from the covariant constancy of $\t e_{(-)}$, 
$$
\t W^{-\,b}_{(-)\,Na}=\t\Omega^{-\,M}_{NK}\t e^K_{(-)\,a}\t e^b_{(-)\,M}
+ \t e^b_{(-)\,M}\del_N\t e^M_{(-)\,a}\,,
$$ 
where $\Omega^{\mp\,M}_{NK}=\G^M_{NK}\mp G^{MP}H_{PNK}$ are the
torsionful connections. Using the $O(d,d)$ transformation of
$\Omega^{-\,M}_{NK}$ as given in \cite{SFH3} and rewriting the result
in terms of $W^{-\,b}_{N\,a}$, one will finally recover (\ref{W--})
after some manipulations. One will also have to use (\ref{Qcompo}) as
well as $Q_+=Q_--(S-R)G$ to simplify $Q^{-1}_-Q_+$. To obtain
(\ref{W++}) one follows the same procedure, but now starting with the
covariant constancy condition for $\t e_{(+)}$.

To obtain (\ref{OddH}) with the lower signs, we start from the
definition of $\t W^-_{(-)Mab}$ which can be written as 
$$
\t e^M_{(-)\,a}\t e^N_{(-)\,b}\t e^K_{(-)\,c} \t H_{MNK}=
2\eta_{ad} (\t w^d_{(-)\,Mc} - \t W^{-d}_{(-)\,Mc})\t e^M_{(-)\,b}\,.
$$
First, we write $\t W^{-}_{(-)}$ in terms of $W^{-}$ and express the
result in terms of $H$. Then, we expand $\t w^d_{(-)\,Mc}$ in terms of
the vielbein $\t e_{(-)}$ and express the result in terms of 
$w^d_{Mc}$ and $W^{-a}_{i\,b}$. After some further manipulations
one gets  (\ref{OddH}) with the lower signs. To get the equation with 
upper signs, replace $e_{(-)}$ by $e_{(+)}$ in the starting
equation. 

\resection{Some ${\bf\Gamma}$-matrix Results}

We use the metric signature $\{-1, +1,\ldots, +1\}$ and define the
$\G$-matrices such that, 
\beq
\{\G^a,\G^b\}=2\eta^{ab} \,,
\label{AppB1}
\eeq
In the Majorana-Weyl representation all $\G^a$ are real, with $\G^0$
antisymmetric and others symmetric. To fuse products of $\G$-matrices
into antisymmetrized ones, we use the identity
\beq
\G_{a_1\cdots a_i}\,\G^{b_1\cdots b_j}=
\sum_{k=|i-j|}^{i+j} \frac{i!\,j!}{s!\,t!\,u!}\,
\delta^{[b_1}_{[a_i}\cdots \delta^{b_s}_{a_{t+1}}\,
\G_{a_1\cdots a_t]}^{{\hphantom{a_1\cdots a_t]}}b_{s+1}\cdots b_j]}\,,
\label{AppB2}
\eeq
with
$$
s=\frac{1}{2}(i+j-k)\,,\quad t=\frac{1}{2}(i-j+k)\,,\quad 
u=\frac{1}{2}(-i+j+k)\,.
$$
In the summation, only those values of $k$ appear for which $s$, $t$
and $u$ are integers, {\it i.e.}, $k=|i-j|, |i-j|+2,\cdots, i+j-2, 
i+j$.
The trace of products of $\G$-matrices is given by
\beq
{\rm Tr}\left(\G^{a_1\cdots a_l}\G_{b_1\cdots b_k}\right)
= 2^5\,\delta_{kl}\, (-1)^{k(k-1)/2}\, k!\,\delta^{[a_1}_{[b_1}
\cdots \delta^{a_k]}_{b_k]}\,. 
\label{AppB3}
\eeq
All antisymmetrizations are with unit weight.


\begin{thebibliography}{99}
\bibitem{KSN}
K. S. Narain, Phys. Lett. {\bf B169} (1986) 41;\\
K. S. Narain, M. H. Sarmadi and  E. Witten, Nucl. Phys. {\bf B279}
(1987) 369.
\bibitem{GRV}
A. Giveon, E. Rabinovici and G. Veneziano, Nucl. Phys {\bf B322} (1989) 167.
\bibitem{MV}
K. A. Meissner and G. Veneziano, Phys. Lett. {\bf B267} (1991) 33. 
\bibitem{SEN}
A. Sen, Phys. Lett. {\bf B271} (1991) 295; Phys. Lett. {\bf B274}
(1992) 34 (hep-th/9108011). 
\bibitem{HS}
S. F. Hassan and A. Sen, Nucl.Phys. B375 (1992) 103-118 (hep-th/9109038).
\bibitem{MS}
J. Maharana and J. H. Schwarz, Nucl. Phys. {\bf B390} (1993) 3 
(hep-th/9207016). 
\bibitem{GR}
A. Giveon and M. Rocek, Nucl. Phys. {\bf B380} (1992) 128 (hep-th/9112070).
\bibitem{BHO}
E. Bergshoeff, C. M. Hull and T. Ortin, Nucl. Phys. {\bf B451} (1995) 547
(hep-th/9504081).
\bibitem{HT}
C. M. Hull and P. K. Townsend, Nucl. Phys. {\bf B438} (1995) 109
(hep-th/9410167).
\bibitem{EW}
E. Witten, Nucl. Phys. {\bf B443} (1995) 85 (hep-th/9503124).
\bibitem{AAFFT}
L. Andrianopoli, R. D'Auria, S. Ferrara, P. Fr\'{e} and M. Trigiante,
Nucl. Phys. {\bf B496} (1997) 617 (hep-th/9611014),\\
L. Andrianopoli, R. D'Auria, S. Ferrara, P. Fr\'{e}, R. Minasian 
and M. Trigiante, Nucl. Phys. {\bf B493} (1997) 249 (hep-th/9612202).
\bibitem{BMZ}
D. Brace, B. Morariu, and B. Zumino, Nucl. Phys. {\bf B549} (1999) 181
(hep-th/9811213).
\bibitem{FOT}
M. Fukuma, T. Oota and H. Tanaka, {\it Comments on T-dualities of
Ramond-Ramond Potentials} (hep-th/9907132).
\bibitem{KS}
A. Konechny and A. Schwarz, Phys. Lett. {\bf B453} (1999) 23
(hep-th/9901077).
\bibitem{SFH4}
S. F. Hassan, {\it T-Duality, Space-time Spinors and R-R Fields in
Curved Backgrounds} (hep-th/9907152).
\bibitem{BS}
I. Bakas, K. Sfetsos, Phys. Lett. {\bf B349} (1995) 448 (hep-th/9502065). 
\bibitem{SFH3}
S. F. Hassan, Nucl. Phys. {\bf B460} (1996) 362 (hep-th/9504148).
\bibitem{SFH2}
S. F. Hassan, Nucl. Phys. {\bf B454} (1995) 86 (hep-th/9408060).
\bibitem{CNLY}
C. G. Callan, C. Lovelace, C. R. Nappi and S. A. Yost, 
Nucl. Phys. {\bf B308} (1988) 221.
\bibitem{POL}
J. Polchinski, {\it TASI Lectures on D-Branes} (hep-th/9611050);\\
J. Polchinski, S. Chaudhuri, C. V. Johnson, {\it Notes on D-Branes}
(hep-th/9602052).
\bibitem{Romans}
L. J. Romans, Phys. Lett. {\bf 169B} (1986) 374.
\bibitem{JHS}
J. H. Schwarz, Nucl. Phys. {\bf B226} (1983) 269.
\bibitem{BMM}
J. C. Breckenridge, G. Michaud and R. C. Myers, Phys. Rev. {\bf D55}
(1997) 6438 (hep-th/9611174).
\bibitem{KHP}
A. D. Kennedy, J. Math. Phys. {\bf 22} (1981) 1330,\\
J. W. van Holten and A. Van Proeyen, J. Phys. {\bf A15} (1982) 3763.
\bibitem{AVP}
A. Van Proeyen, {\it Tools for Supersymmetry} (hep-th/9910030).
\bibitem{BRGPT}
E. Bergshoeff, M. de Roo, M. B. Green, G. Papadopoulos and P. K. Townsend, 
Nucl. Phys. {\bf B470} (1996) 113 (hep-th/9601150).
\bibitem{CLPS}   
M. Cveti\v{c}, H. L\"{u}, C. N. Pope and K. S. Stelle, {\it T-duality
in the Green-Schwarz Formalism and the Massless/Massive IIA Duality
Map}, (hep-th/9907202).   
\bibitem{PS}
B. Pioline and A. Schwarz, JHEP 9908 (1999) 021, (hep-th/9908019).
\bibitem{DMWY}
A. Dhar, G. Mandal, S. R. Wadia and K. P. Yogendran, {\it $D1/D5$
System with B-Field, Noncommutative Geometry and the CFT of the Higgs
Branch}, (hep-th/9910194). 
\end{thebibliography}
\end{document}